\documentclass[sigconf]{acmart}

\setcopyright{none} 

\usepackage{color}
\usepackage{ifthen}
\usepackage{amssymb}
\usepackage{url}
\usepackage{graphicx}
\usepackage{listings}
\usepackage{lmodern}

\usepackage{framed}
\usepackage[T1]{fontenc}
\usepackage{inputenc}
\usepackage{multirow}
\usepackage{multicol}
\usepackage{tabularx}
\usepackage{booktabs}
\usepackage{algorithm}
\usepackage{algorithmicx}
\usepackage{algpseudocode}
\usepackage{enumitem}
\usepackage{makecell}

\definecolor{codeblue}{RGB}{0,0,128}
\definecolor{codegreen}{RGB}{0,128,0}
\definecolor{codefullblue}{RGB}{0,0,255}

\lstdefinestyle{ruleStyle}{
	keywordstyle=\color{codeblue},
	keepspaces=true,
	breaklines=true,
	stringstyle=\color{codegreen},
	numberstyle=\color{codegreen},
	basicstyle=\small
}

\lstdefinelanguage{ruleLge}{
	morekeywords={rule, when,then,end},
	sensitive=false,
	morestring=[b]"
}

\lstdefinelanguage{ruleMeta}{
	morekeywords={metamodel, ruleDef, STRING, condition, action, term, boolOperator, attribute, NUMBER, type, IDENT, task, operation, value, op},
	sensitive=true,
	morestring=[b]'
}

\newboolean{showcomments}
\setboolean{showcomments}{true}
\ifthenelse{\boolean{showcomments}}
 { \newcommand{\mynote}[2]{
      \fbox{\bfseries\sffamily\scriptsize#1}
        {\small$\blacktriangleright$\textsf{\textcolor{red}{{\em #2}\bf }}$\blacktriangleleft$}}}
        { \newcommand{\mynote}[2]{}}

\newboolean{hidetext}
\setboolean{hidetext}{false}
\ifthenelse{\boolean{hidetext}}
  { \newcommand{\hide}[1]{}}
  { \newcommand{\hide}[1]{#1}}

\newcommand{\etal}{\textit{et al., }}
\newcommand{\eg}{\textit{e.g., }}
\newcommand{\ie}{\textit{i.e., }}
\newcommand{\cf}{\textit{cf. }}

\DeclareTextSymbol{\degre}{OT1}{23}

\acmConference[MOMO'17]{Modularity in Modelling Workshop}{April 2017}{Brussels, Belgium}

\begin{document}

\title[Weaving Rules into Models@run.time for Embedded Smart Systems]{Weaving Rules into Models@run.time \texorpdfstring{\\}{} for Embedded Smart Systems}

\author{Ludovic Mouline}
\author{Thomas Hartmann}
\author{Francois Fouquet}
\author{Yves Le Traon}
\affiliation{University of Luxembourg}
\email{firstName.lastName@uni.lu}

\author{Johann Bourcier}
\author{Olivier Barais}
\affiliation{Universite de Rennes 1}
\email{firstName.lastName@irisa.fr}

\renewcommand{\shortauthors}{L. Mouline et al.}

\begin{abstract}
Smart systems are characterised by their ability to analyse measured data in live and to react to changes according to expert rules.
Therefore, such systems exploit appropriate data models together with actions, triggered by domain-related conditions.
The challenge at hand is that smart systems usually need to process thousands of updates to detect which rules need to be triggered, often even on restricted hardware like a Raspberry Pi. 
Despite various approaches have been investigated to efficiently check conditions on data models, they either assume to fit into main memory or rely on high latency persistence storage systems that severely damage the reactivity of smart systems.
To tackle this challenge, we propose a novel composition process, which weaves executable rules into a data model with lazy loading abilities.
We quantitatively show, on a smart building case study, that our approach can handle, at low latency, big sets of rules on top of large-scale data models on restricted hardware.
\end{abstract}

\keywords{Models@run.time, Reactive systems, Rule engines, Lazy loading, Smart systems, Embedded Systems}

\maketitle

\section{Introduction and Motivation}
\label{sec:intro}

To make sustainable decisions and to take appropriate actions, smart systems need to continuously analyse their context, \ie their environment and internal state~\cite{DBLP:conf/seke/0001FNMKT14, DBLP:conf/models/0001FNMKBT14}. 
For example, a smart building can contain hundreds of devices that continuously generate data which contribute to an understanding of the context.
The interest of end users with such systems is, for example, to know the current state of the building, \eg the current temperature, or to remotely switch on/off a heating system.
To do so, such systems often need to process thousands of data updates---\eg sensor values---per second in near-time and on hardware with limited computational capabilities, like a Raspberry Pi~\citep{vujovic2015raspberry}.
Processing these updates mostly consists in a verification against a set of domain-defined conditions that can trigger specific actions~\citep{Wu:2006:HCE:1142473.1142520}. 
Which actions need to be triggered for which updates, \ie for which \textit{patterns}, can be defined in so-called \textit{rules}. 
Performance of processing these rules is key critical in such domains, since it directly defines the reactivity level of a smart system.
This applies even more when considering security rules that need to aggregate various data to fire the relevant counter-actions.
For instance, in smart building, various temperature sensors (\eg indoor and outdoor) need to be correlated to detect that a door is open and to ultimately fire an alarm.
Thus, independent time series are not suitable to correlate data from different data sources. 
Instead, efficiently correlating data relies on navigable data structures~\citep{DBLP:conf/seke/0001FNMKT14}.
The models@run.time paradigm~\citep{DBLP:journals/computer/MorinBJFS09, DBLP:journals/computer/BlairBF09} has proven its suitability to represent the context of such systems and to provide a navigable structure for reasoning engines.
To accurately reflect a current system context, models@run.time are regularly updated, \eg with sensor measurements.
Smart systems need mechanisms to process these updates and trigger actions based on pattern detection---for many domains, in near real-time.

Several approaches try to address the challenge of live processing updates to detect which actions need to be triggered. 
\textit{Complex Event Processing} (CEP)~\citep{luckham2008power} investigates how to detect predefined \textit{events}, \ie particular patterns, like sequences of specific values.
Others suggest to use rule engines, like Drools \citep{browne2009jboss}.
In the home automation domain, rule engines like IFTT~\citep{IFTT}, openHab~\citep{openhab} and Pimatic~\citep{pimatic} are used to automate actions. 
For patterns on complex structures, OCL-like queries have been defined on top of MOF-based models~\cite{DBLP:conf/seke/AvilaSCY10}.
All these approaches, model-based or not, rely on that rules and data models fit completely into main memory or on high latency persistence storages, which severely damage the reactivity of smart systems.
This leads to limitations for systems, which need to process large amounts of rules on limited hardware.
To address these limitations, we propose to combine a set of \textit{if \textless pattern(context)\textgreater then \textless actions\textgreater} within models@run.time structure with lazy loading abilities.
By combining lazy loading mechanisms with a low-latency persistent storages, we do not assume that rules or models must fit completely into main memory.
More specifically we investigate the following research questions:
\begin{itemize}
  \item \textbf{RQ1}: How can we process rules, on limited hardware, with nearly constant memory, regardless of the model size?
  \item \textbf{RQ2}: Despite the lazy loading mechanism, can we obtain sufficient latency to enable near real-time process?
\end{itemize}

The remainder of this paper is as follows. 
Section~\ref{sec:contribution} describes our contribution, Section~\ref{sec:evaluation} its evaluation, Section~\ref{sec:rw} discusses related work and Section~\ref{sec:conclusion} concludes the paper.
Background is explained in the sections when needed.

\section{Weaving Rules into Models@run.time}
\label{sec:contribution}
In this section we detail our rule language and weaving process to combine rules and models@run.time.

\subsection{Language Definition}
\label{sec:language}

To weave rules into models@run.time, we leverage two kinds of input: one for the definition of the data structure and another one for defining rules.
For the data structure, we reuse common meta-model formalisms, such as defined by MOF and implemented by EMF/Ecore~\citep{budinsky2004eclipse}.
A data model is a set of classes, which contain a set of attributes and references to other classes.
To meet models@run.time requirements, we use the Kevoree Modeling Framework (KMF)~\citep{DBLP:conf/models/FouquetNMDBPJ12, DBLP:journals/corr/FrancoisNMDBPJ14}, which has been specifically designed for this purpose.
Using a textual syntax, KMF allows to define data structures with built-in lazy loading mechanisms, used for this approach in the processing engine (\cf Section~\ref{sec:processing}).
For the second kind of input, rule definitions, we reuse state-of-art rules modelling concepts, where each rule is composed of: a \textit{name}, a \textit{condition}, and an \textit{action}.
The grammar of our rule modelling language is depicted in Listing~\ref{code:rules-grammar}.
This language is inspired by the \textit{when \textless condition\textgreater then \textless actions\textgreater} pattern.

\begin{figure}
\begin{lstlisting}[language=ruleMeta,caption=ANTLR grammar of the rules language, label=code:rules-grammar, basicstyle=\scriptsize]
metamodel: ruleDef*;
ruleDef: 'rule' STRING condition action 'end';
condition: 'when' ('not')? (term op term);
term: (type '.' attribute) | NUMBER | STRING;
op: ( '==' | '>' | '>=' | '<' | '<=' | '!=');	
type: IDENT ('.' IDENT)*;
attribute: IDENT;
action: 'then' task;
task: operation ('.' operation)*;
operation: IDENT '(' (value(',' value)*)? ')';
value: STRING | '{' task '}';
\end{lstlisting}
\end{figure}

Conditions allow to specify two things: \textit{i)} to which class a rule is attached, we will refer to this as the context of the rule, and \textit{ii)} the condition of the context to trigger the execution of an action, \ie a condition on values.
The current version of our rule language can only define rules based on single attributes.
We intent to extend this in future work, however this is out of scope of this paper, which focus on the performance and lazy-loading impact of rule engines.
For rule action definition, we reuse the formalism proposed by Gremlin~\citep{gremlingithub}, which has proven its expressiviness to define a flow of processing actions on graph structures.
Our rule action can be regarded as a pipeline, where each action is chained and propagates results to the next one.
We provide actions for four kinds of operations: \textit{i)} to navigate in the model, \textit{ii)} to manipulate the control flow, \eg an \textit{if} statement, \textit{iii)} to manipulate the result, \eg a filter, \textit{iv)} to manipulate variables, like saving results in a variable.
Currently, all rules have its own graph condition.
In future work we plan to use approaches like Rete~\citep{forgy1982rete}, which suggest to share condition trees.

An example rule is shown in Listing~\ref{code:rule-example}, relying on a simple meta-model where a class \textit{Room} has a relationship to a class \textit{HeatingSystem}, containing an attribute \textit{status}.
The then block considers the current room as the starting point to chain actions, such as \textit{traverse}.
As effect, this simple rule activates the heating system when the temperature is below 18 degree.
To trigger actions, such rules just modify the current model itself by expecting a synchronisation by a models@run.time engine.
In addition we support arbitrary action code through injection of lambda functions within the chain of actions.

\begin{lstlisting}[language=ruleLge, caption=Rule example, label=code:rule-example]
rule "SwitchOnHeatingSystem"
when
	building.Room.temperature < 18
then
	relation('heatingSystem')
	.setAttribute("status","on")
end	
\end{lstlisting}


\subsection{Rule Action Compilation Process}
\label{sec:compilation_process}

\begin{figure}
	\centering
	\includegraphics[width=0.4\textwidth]{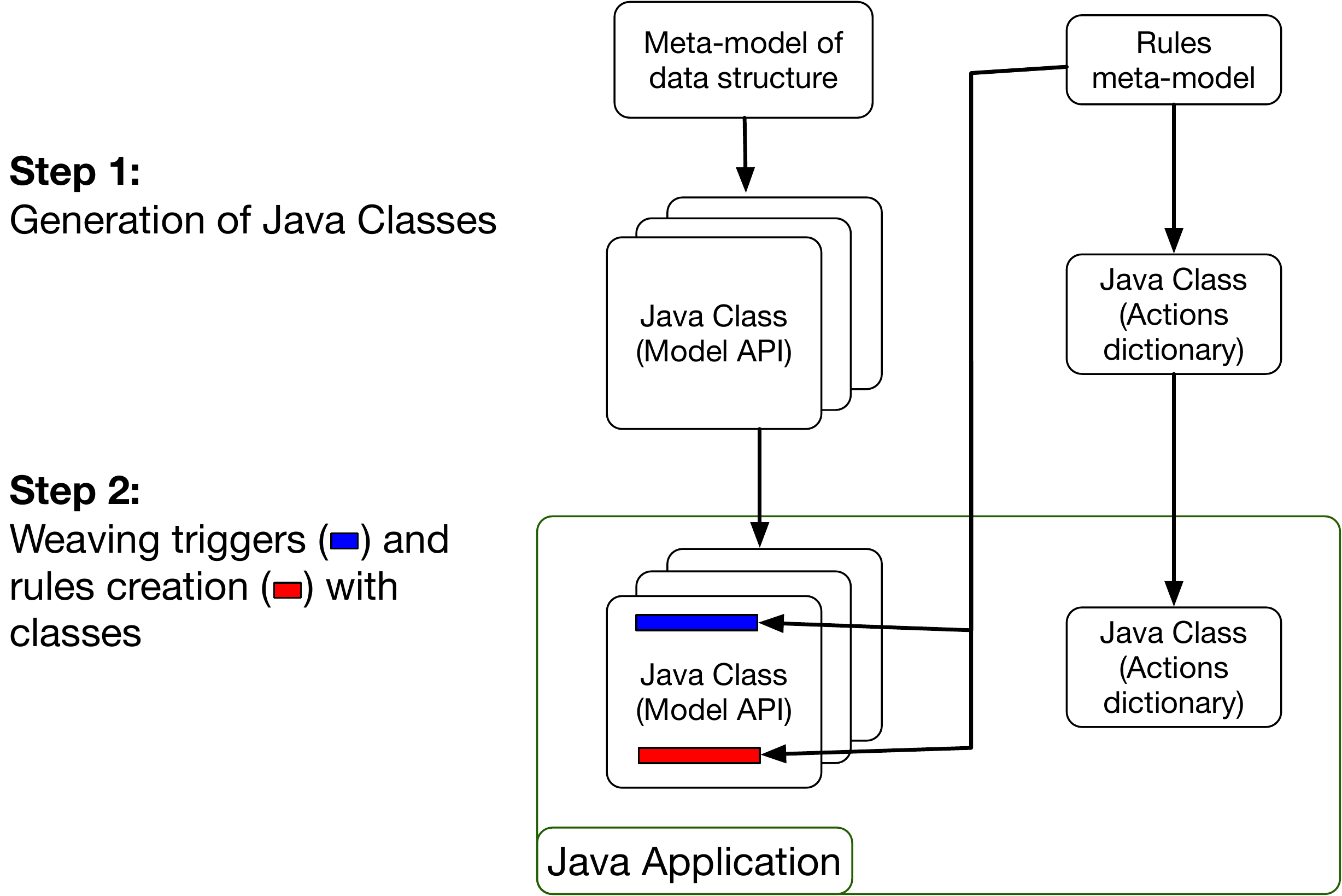}
	\caption{Compilation process}
	\label{fig:compilation-process}
\end{figure}

In this section, we describe the weaving process to inject executable rules during the generation process of KMF.
The result of this weaving process is a standalone artefact, which  contains Java classes ready to be used as a models@run.time backbone.
An overview of this is shown in Figure~\ref{fig:compilation-process}.

KMF generates, based on a meta-model, a set of classes, which are referred to as \textit{model API}.
This API allows to manipulate data compliant to a meta-model during execution: creating new model elements, modifying existing ones, or navigating inside the model.
Every KMF model element is lazily loaded during the execution when relationships are traversed.
Once elements are modified in-memory, KMF persists them to disk using KeyValue storages, such as RocksDB.
To automatically trigger the verification of rules, setter methods of the KMF API are overriden for classes concerned by any rule.
This generation process can be seen on the upper-left part of Figure~\ref{fig:compilation-process} and is divided in two steps.
The first one, depicted on the upper-right part of Figure~\ref{fig:compilation-process}, consists in the creation of one Java class containing all actions, called \textit{Actions dictionary}.
Actions are compiled into static Java code, identified within the dictionary with an integer value.
This way, when a rule condition triggers an action, the system can simply hit the dictionary with a previously stored integer reference to execute the corresponding action code.
The second step aims at generating the trigger code to verify automatically rules.
Because conditions are always based on model element value updates, we override the model API setter to trigger all rule conditions related to this particular class and attribute.
This is depicted as blue rectangle in Figure~\ref{fig:compilation-process}.
As a result, we obtain a standalone Java artefact, embedded all actions as static methods automatically called when KMF elements are modified through an extended setter.
This, together with the KMF lazy loading mechanism, allows to workaround the need to keep every model element in memory to listen for updates.

\subsection{Weaving Condition Trees and Models}
\label{sec:weaving}

\begin{figure}
	\centering
	\includegraphics[width=\linewidth]{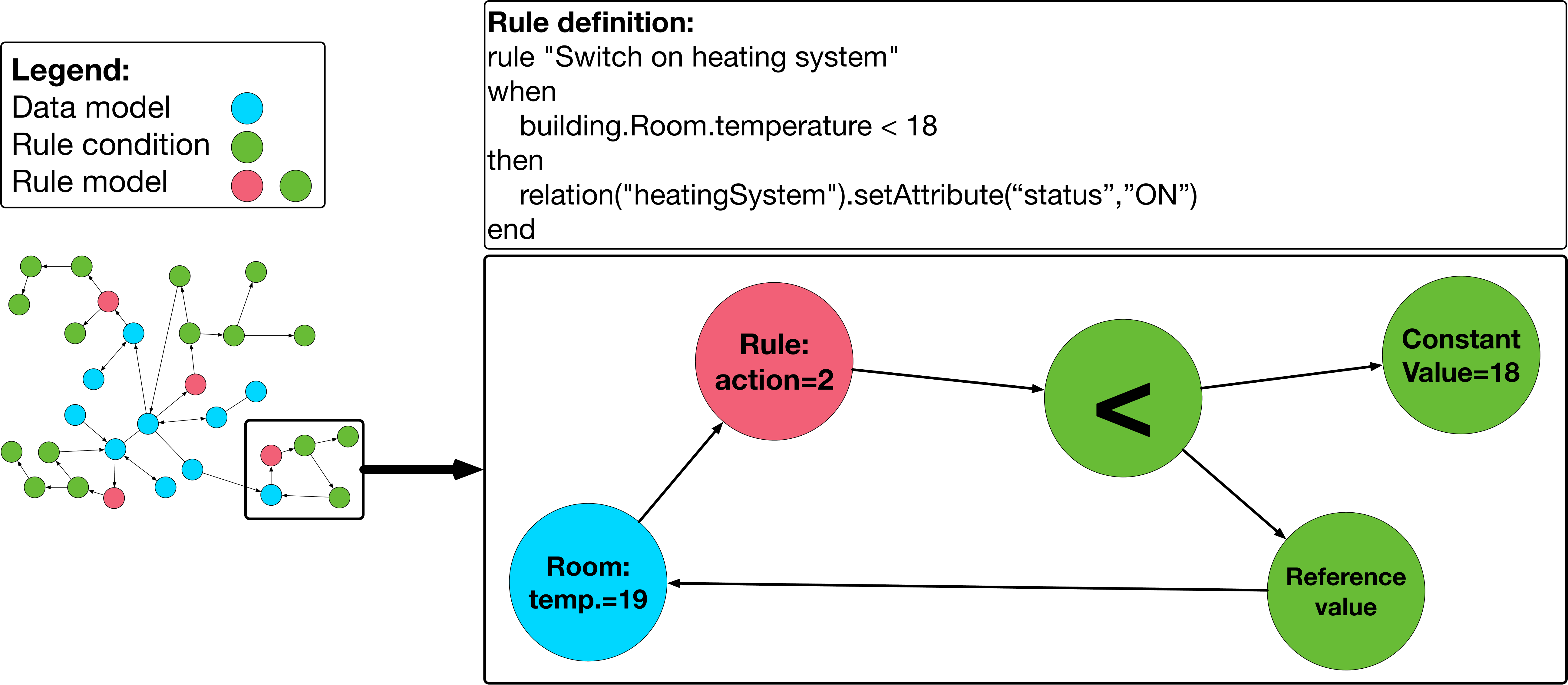}
\caption{Model of a rule condition}
\label{fig:ast-condition}
\end{figure}

In this Section, we describe how rule conditions are weaved into the model.
Like most object-oriented modelling frameworks, KMF uses a graph-based approach to model a system, \ie the model can be seen as a graph of interacting objects, where the graph structure conforms to the meta-model.
Each node in the graph conforms to a meta-class and is editable and accessible using the model API, generated as explained in Section~\ref{sec:compilation_process}.
The node related to the KMF classes are referred to as data node.
These nodes are depicted in blue in Figure~\ref{fig:ast-condition}.
For rules, we use two parts: a \textit{rule node} and a \textit{condition graph}.
A rule node represents a rule, stores actions, and has a relationship to the graph condition.
As mentioned in the previous section, actions are compiled into static fields, using integers as identifiers.
We store this identifier as an attribute in a rule node.
Rule nodes are depicted in red in Figure~\ref{fig:ast-condition}.

Moreover, in Section~\ref{sec:language}, we explained that a rule first defines a context, \ie the class for which the rule should be applied.
A rule node is created for each data node, implied in a rule as context and that conformes to a KMF class.
These two nodes are linked, to first enable an efficient rule verification and secondly to set the input of the first operation with the data node.
Additionally, rule nodes contain an integer reference to an executable action present in the dictionary.

Rule conditions are modelled using a graph, which represents its \textit{Abstract Syntax Tree} (AST).
For each different term and operator, there is a specific type of nodes.
Currently, we defined twelve different nodes for this AST---not all of them are currently used in the rule language, \ie statements for the rule condition: boolean operator (and, or, not), arithmetic operator(=,!=,\textgreater,\textgreater =,\textless, \textless =), constant value and reference value, that refers to the value of another node.
The AST can be modelled using these nodes.
An example is depicted in green in Figure~\ref{fig:ast-condition}.
On the upper-right part of  the figure, in the rectangle, the result of the compilation of the rule is shown in Listing~\ref{code:rule-example}.
On this graph, we can see the second composition of rule nodes and data model nodes: the condition node \textit{Reference value} has a relationship with the node \textit{Room}, which is part of the data model.

\subsection{Rule Processing}
\label{sec:processing}

\begin{figure}
	\centering
	\includegraphics[width=\linewidth]{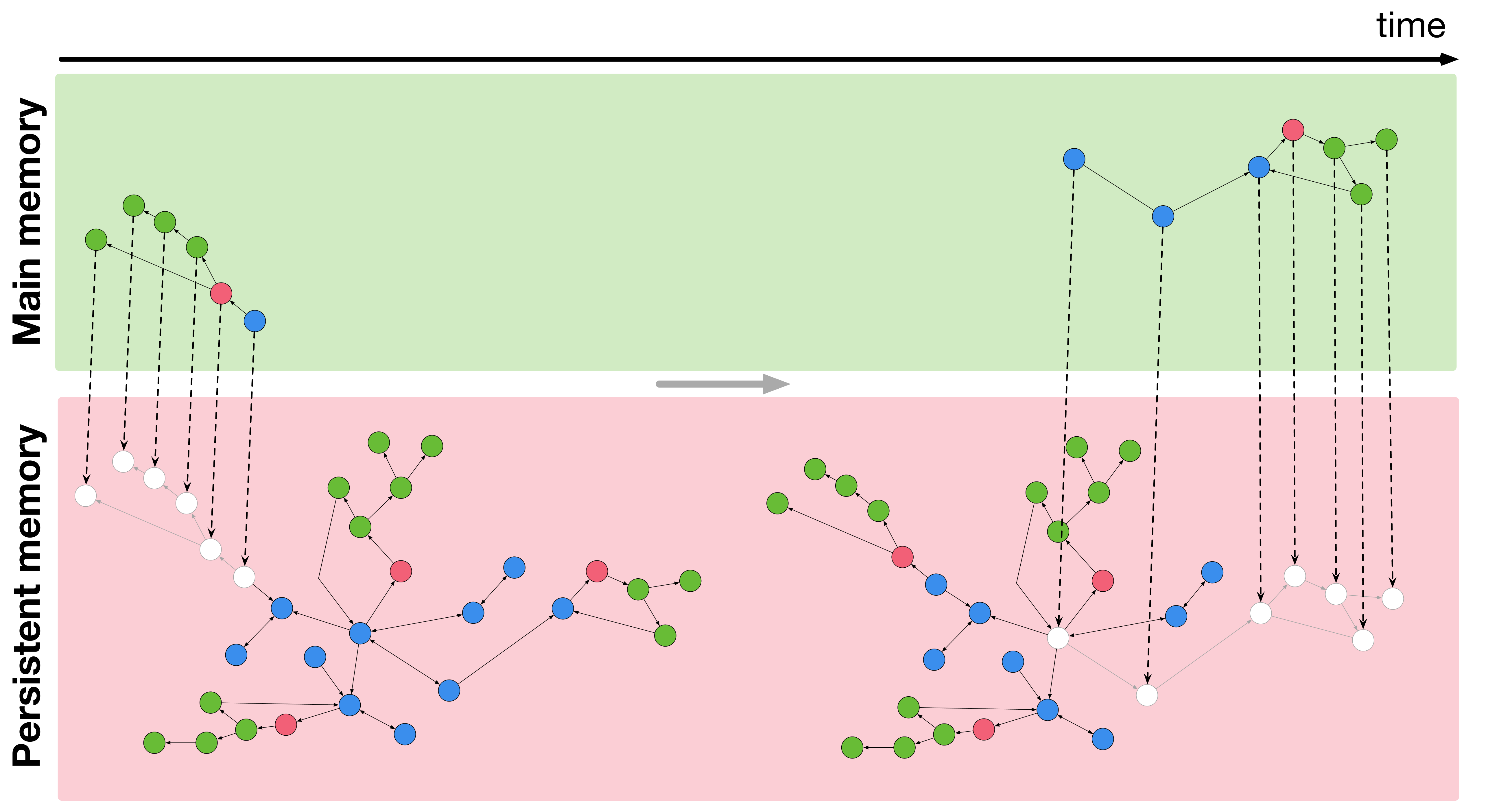}
	\caption{Lazy loading during graph navigation}
	\label{fig:lazy-loading}
\end{figure}

In this section, we describe how we traverse the model, woven from the data structure definition and rule definition, using lazy loading techniques.
When an attribute, which is part of a rule definition, is modified, we navigate through the relations of a rule, process the condition graph and, if the condition is validated, get the action using its identifier and executed it.
This process uses lazy loading techniques to dynamically load the necessary node on demand into main memory~\citep{DBLP:conf/models/0001MFNKT15}.
When a data model node is accessed, the system first looks into the main memory if the node is present.
If not, it will load it from a persistent storage, and vice versa, stores unused nodes if needed in order to free memory. 
Figure~\ref{fig:lazy-loading} shows the processing of two rules, with a memory size that can contain six elements.
The upper zone describes model elements that are loaded in main memory, while the lower zone shows a view on a composed graph.

To process the condition graph AST, we leverage a classical binary tree interpretation.
Each node involved in the condition AST has at most two children.
If a node has at least one children, it means that it is an operator.
There is only one exception: the \textit{node value} that returns the value of an attribute of a data node.
A node value is defined by two things: a relation to the target node and the name of the attribute stored as an attribute. 
For instance, in the example depicted in Figure~\ref{fig:ast-condition}, the reference node has a relationship to a room node and store the attribute name 'temperature' as its own attribute. 
Processing such nodes mean: resolve the node of the relationship and return the value stored in the attribute, in our example the temperature attribute.
To compute the value of an operator node, the system needs to get the values of its children and apply the semantic of the operator.
For nodes without children, the node returns the stored value or applies a process to compute or get one.

\section{Evaluation}	
\label{sec:evaluation}
%
%
%
%

To answer the two research questions formulated in Section~\ref{sec:intro}, we conducted in this section an experimental evaluation.
Source code is publicly available on GitHub\footnote{https://github.com/lmouline/momo17-bench}.
For all experiments we rely on a smart building dataset with simulated sensor values.

\begin{figure*}
	\includegraphics[width=\linewidth]{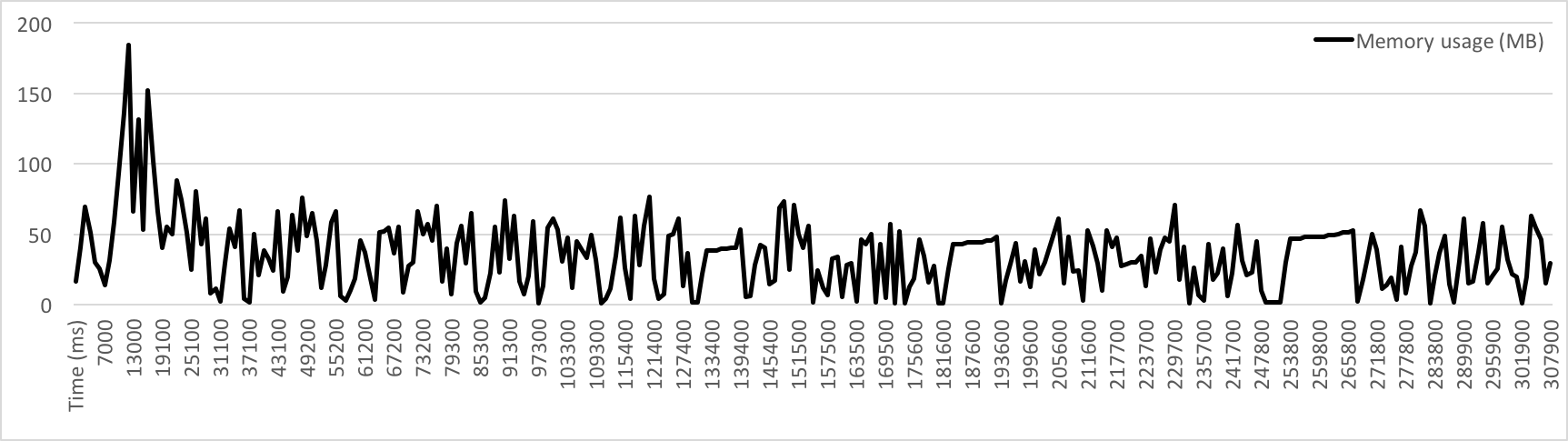}
	\caption{Results for RQ1}
	\label{fig:res-exp-rq1}
\end{figure*}

\paragraph{RQ1: processing rules with constant memory}
In our approach, we defined a combination of model and condition AST with lazy loading abilities.
As a result, we should be able to run with an arbitrary size of memory, regardless of the actual model size.
If rules are sequentially processed, the memory limitation is givem by the number and size of nodes implied in one rule (condition AST, rule node, and related data node).
In case rules are processed in parallel, the memory requirement is to fit at least the rule with the largest number of nodes implied for all  threads.
During this experiment, we fix the cache size of KMF to 10,000 elements to force the lazy loading mechanism to mainly work from disk.
Moreover, the model size increases from 100,000 elements to 5,000,000 elements.
During each iteration, we sequentially check all rule conditions and trigger those that are evaluated to true.
Figure~\ref{fig:res-exp-rq1} shows that memory consumption stays constant and below 50MB all along the process.
From these results we can conclude that our approach allows to process a massive model even with limited memory.

\paragraph{RQ2: latency of rule checking using lazy loading} 
Because our approach relies on a lazy loading mechanism with a persistence storage, we drastically reduce the memory usage, as shown in RQ1.
However, these benefits come at the price of potentially decreasing the latency of the rule engine.
We conducted a second experiment to quantify the latency of our approach against various rule sizes and numbers.
We setup a model of 1 million elements (simulating sensor values) and a RocksDB storage.
Then, we measured three batches of verification with rules, containing small to large conditions to check.
For every batch we evaluated 100,000 rules and measured the total execution time: time to load nodes, process the AST condition and execute the action.
As there is a direct link between the execution time of an action and the process time of a rule, we do not modify this parameter in our study.
Results are presented in Table~\ref{table:ts1} in rules per second processed.
From these results we can conclude that despite the lazy loading mechanism and less than 200MB memory allocated for the processing JVM the throughput is still above 50,000 rules per seconds in average.
Such an engine can be embedded in devices like a Raspberry Pi, with less than 500MB memory.

\begin{table}
	\caption{Processing throughput (rules/seconds)}
	\vspace{2mm}
	\label{table:ts1}
	\begin{center}
		\begin{tabular}{ c | c  }
			\Xcline{1-2}{0.7pt}
			\textbf{Condition rule size} & \textbf{Throughput} \\
			\Xcline{1-2}{0.7pt}			
			3 & 70,028 \\
			\Xcline{1-2}{0.7pt}
			31 & 58,788 \\
			\Xcline{1-2}{0.7pt}
			255 & 41,152  \\
			\Xcline{1-2}{0.7pt}			
		\end{tabular}
	\end{center}
\end{table}

\section{Related Work}
\label{sec:rw}
The execution of rules on top of models has been previously discussed by the modelling community.
For example, to handle model to model or model to text transformations.  
In~\citep{varro2002designing}, Varr\'{o} \etal define rules as a set of three elements: a graph pattern to look for, a set of application conditions, and a graph result.
Bergmann \etal~\citep{bergmann2010incremental} proposed EMF-IncQuery, a model transformation framework, based on graph pattern matching, for big models.
Other solutions which have been suggested in the context of model transformations, are the ATLAS Transformation Language (ATL)~\citep{jouault2008atl},  Henshin~\citep{biermann2008precise}, and Jouault \etal~\citep{DBLP:conf/icmt/JouaultT10}.
To enable live processing of rules,~\citep{david2014streaming} analyses model modifications using a CEP engine. 
These approaches require that all data and rules are fully in-memory, whereas our solution loads only the currently processed elements on demand into main memory. 
Furthermore, for these approaches, all rules are stored aside of the model, whereas our solution suggests to combine model and rules.
Textual OCL~\citep{warmer2003object} is also related to our approach.
It allows to define model constraints and derived attributes.
Approaches \citep{avila2010runtime} have been investigated to check model constraints during the execution of a system, \eg in~\citep{DBLP:conf/serp/AvilaFC08}, the authors propose a solution to generate Java Modelling Language (JML)~\citep{DBLP:journals/sigsoft/LeavensBR06} assertions, a language to specify pre and post conditions on top of Java methods, from OCL.
These approaches are made for model checking, whereas we propose a solution for rules, \ie they do no support to execute actions.
Another approaches consist in providing OCL interpreter \citep{DBLP:conf/uml/RichtersG00}.
This approach executes the constraints on model snapshots, which are regularly taken and cannot process the events in a short amount of time.
Another application for rules is goal modelling, where goal models represent goals and scenarios of a system with languages like URN~\citep{urn}. 
~\citep{DBLP:journals/eceasst/Robinson08} and~\citep{vrbaski2012goal} define an approach to combine rule engines with goal modelling techniques.
These approaches rely on an external rule engine.
To efficiently query large models, several approaches have been investigated.
EMF-Query \citep{emfquery} defines an API to access model elements.
To address its difficulty to deal with large models, they propose a new version, EMF-Query 2, that can lazy load the model element from persistent memory.
Recently, a new approach has been proposed to query large models efficiently: the Mogwa\"i framework \citep{DBLP:conf/rcis/DanielSC16}. 
OCL constraints are compiled to Gremlin \citep{gremlingithub} and then executed at the database level. 
This approach allow to generate a Gremlin request from OCL constraints.
Contrary to our approach where our rules are directly executed at the main memory level, this approach implies that the request are executed at the database level.
Furthermore, the approach defends in \citep{gremlingithub} and the one explains in \citep{emfquery} have only been designed to query large models, thus cannot execute actions. 


\section{Conclusion and Perspective}
\label{sec:conclusion}

%
	
Executing rules or queries on large data models is still an open challenge in the modelling community, especially when executed on limited hardware. 
Solutions proposed so far, mostly rely on loading the full model and all rules in memory or push the problem to the database layer, resulting in high latency.
In this paper we presented a novel approach to weave rules into models@run.time using a lazy loading mechanism, in order to deal with the execution of rules on large-scale models.
We claim that most rules do not require the full data model but only relatively small parts.
Therefore, we first proposed to weave rule models into data models. 
Secondly, we used a lazy loading mechanism to load/unload required data on demand.
We integrated this approach into the Kevoree Modeling Framework and showed that it can handle thousands of rules, combined in a large model, with a small and constant memory consumption.

We plan to extend this approach in future work on several aspects.
Firstly, we want to provide a more expressive language to define rules and manage conditions that are based on more than one attribute.
Our approach is restricted to one attribute.
To extend it to several attributes, we intend to explore a publish-subscribe-based system combined with a buffer-based system.
The Pub-Sub system should be able to notify a rule node that a specified attribute has changed.
The buffer-based system should allow to synchronise different updates coming from several attributes, belonging to one or different nodes.
Secondly, we aim at improving the memory required to store the model by using a Rete like approach to represent conditions.
And finally, we want to introduce temporal aspects in rules.

\begin{acks}
This work was supported by POST Luxembourg.		
\end{acks}


\end{document}